\documentclass[12pt]{iopart}
\usepackage{latexsym}

\begin{document}

\title[\bf Quantum Preparation]{\bf
A Theory of Quantum Preparation and the
Corresponding Advantage
 of the Relative-Collapse Interpretation
 of Quantum Mechanics
 as Compared to the Conventional One}
\author{F Herbut\footnote[1]{E-mail:
fedorh@infosky.net}}
\address{Faculty of Physics, University of
Belgrade, POB 368, Belgrade 11001,
Yugoslavia and Serbian Academy of
Sciences and Arts, Knez Mihajlova 35,
11000 Belgrade}

\date{\today}
-----------------------------------------------------------------\\
\noindent Analyzing two standard preparators, the Stern-Gerlach and the
hole-in-the-screen ones, it is demonstrated that four entities are
the basic ingredients of the theory: the composite-system
preparator-plus-object state (coming about as a result of a
suitable interaction between the subsystems), a suitable
preparator observable, one of its characteristic projectors called
the triggering event, and, finally, the conditional object state
corresponding to the occurrence of the triggering event. The
concepts of a conditional state and of retrospective apparent
ideal occurrence are discussed in the conventional interpretation
of quantum mechanics. In the general theory of a preparator in
this interpretation first-kind and second-kind preparators are
distinguished. They are described by the same entities in the same
way, but in terms of different physical mechanisms. In this
article the relative-collapse interpretation is extended to
encompass also preparators (besides measuring apparatuses). In
this interpretation also the mechanisms become the same and one
has only one kind of preparators. \\
--------------------------------------------------------------

\section{INTRODUCTION}

\rm \large \noindent In Sections 2-6 we discuss a theory of quantum preparators in the
context of usual quantum mechanics. In Section 7 we enter the
relative-collapse interpretation of quantum mechanics and show
that the expounded preparator theory is simpler and more natural
in this interpretation.
\par  We start by constructing a preparator out of a measuring device.
The most elementary of the latter is the Stern-Gerlach (S-G)
spin-projection measuring instrument described in almost every
textbook on quantum mechanics.

\section{STERN-GERLACH PREPARATORS}

\indent  To begin with, let us sum up some of the familiar  features
of the S-G measuring instrument in order to single out the
relevant ones important both in the conventional and in the
relative-collapse interpretations of quantum mechanics.
\par We assume that it is the z-projection of the spin that is  measured.
As well known, the magnetic field couples the spatial degrees of
freedom of the outgoing particle (leaving the field and
approaching the plates) with its z-projection of spin as follows:
$$ |\Phi \rangle _{12} \equiv \alpha |\psi ^+ \rangle _1|+,z \rangle _2 + \beta |\psi ^- \rangle _1|-,z \rangle _2 \eqno{(1)}$$
if the incoming particle was in the uncorrelated pure state given
by the state vector
$$ |\Psi \rangle _{12} \equiv |\psi^0 \rangle _1 \Big( \alpha |+,z \rangle _2 + \beta |-,z \rangle _2 \Big) .$$
Here $\alpha ,\beta \in \mbox{\bf C}, \quad |\alpha |^2 +
|\beta|^2=1$; the first subsystem consists of the spatial degrees
of freedom of the particle, and the second one is that of spin;
$|\psi ^+ \rangle _1,\, |\psi ^- \rangle _1,$\, and $|\psi ^0 \rangle _1 $ are the outgoing
upward moving, the outgoing downward moving and the incoming
spatial state vectors respectively; and, finally, $|\pm ,z \rangle _2$ are
the spin-up and spin-down (along z) state vectors respectively.
\par  Let us introduce the projectors

$$ P^+_1 \equiv \int ^{+\infty}_{-\infty} \int ^{+\infty}_{-\infty} \int ^{+\infty}_0 |x,y,z \rangle _1 \langle x,y,z|_1dxdydz, \eqno{(2a)} $$

$$ P^-_1 \equiv \int ^{+\infty}_{-\infty} \int ^{+\infty}_{-\infty} \int ^0_{-\infty }|x,y,z \rangle _1 \langle x,y,z|_1dxdydz   \eqno{(2b)} $$
projecting onto the upper and the lower halfspace respectively. We
define
$$  A_1 \equiv a_+P^+_1 + a_-P^-_1 \eqno{(3)} $$
with arbitrary but fixed $a_+\not= a_-, \quad a_+,a_-\in
\mbox{\bf R}$.
\par Thus we have obtained  the four basic entities  for  our  theory  of
the preparator (in both interpretations):
$$ |\Phi \rangle _{12}, \quad
A_1, \quad P_1^{(\bar n )}\equiv P^+_1, \quad  \rho _2^{(\bar n )}
\equiv |+,z \rangle _2 \langle +,z|_2, \eqno{(4a,b,c,d)} $$
where $"\bar n "$ is the quantum number fixing a particular
characteristic event (projector) out of those appearing in the
spectral form (3). We call the singled-out event \it the
triggering event \rm of preparation, and we take into account that

$$ P^+_1 |\psi ^+ \rangle _1 = |\psi ^+ \rangle _1, \quad P^+_1|\psi ^- \rangle _1=0. $$
Hence, the ideal occurrence of the first-subsystem event
$(P^+_1\otimes 1)$ in the composite-system state $|\Phi \rangle _{12} $
brings the second subsystem into the state $|+,z \rangle _2$. (According
to the L\"{u}ders $\mbox{\rm formula}^{(1)}$ -cf also
$\mbox{\rm Messiah}^{(2)}$ -one applies the projector onto
the state vector and one renormalizes the result.)
\par It is noteworthy that the composite-system state (4a)
and the first-subsystem observable (4b) with a purely discrete
spectrum are completely independent of each other. The particular
characteristic projector (4c) is very much related to the
mentioned observable, and the second-subsystem state (4d) is
related to the mentioned projector as the state which comes about
when the event represented by the projector occurs in the
composite system state (4a). (More about this in Section 4.)
\par The S-G measuring apparatus performs  nonrepeatable  or  second-kind
measurement in its standard form (when the particles are stopped
on the plates). Therefore, to obtain a preparator, some
modification is required.

\subsection{The First Modification for a S-G Preparator}

\indent \rm We assume that the S-G device is modified so that the upper plate is
removed, and in its place we have a detector that detects the
presence of the particle, but so that
\begin{quotation}
(i) the particle is not stopped; it leaves the device, and
\end{quotation}
\begin{quotation}
(ii) the detector does not interact with the particle by
electromagnetic interaction.
\end{quotation}
\par  Evidently, requirement (i) is necessary to  have  a  first-kind
(or repeatable) measurement, the only kind that may amount to a
preparation. Requirement (ii) is indispensable because we need a
measurement of the first-subsystem observable $A_1$ , i.e., of
$(A_1\otimes 1)$, in the state $|\Phi \rangle _{12}$ (to avoid a spin
flip on the particle due to the absorption of a photon).
\par Once the particle has left the magnetic field  of  the  S-G  device,
there is nothing to couple the spatial and the spin degrees of
freedom (the two subsystems do not interact) in the time interval
$t_i\leq t\leq t_f$. Here $t_i$ is the instant when the
interaction in the (1+2)-system ends, the state $|\Phi \rangle  _{12}$ is established, and the (instantaneous) triggering event occurs in $|\phi _{12} \rangle $. At $t_i$ the preparation is completed. It is the initial moment of the quantum experiment.
We denote by $t_f$ the subsequent instant when an (instantaneous)
measurement on the second subsystem is performed and the final
moment of the experiment is thus reached.
\par It is important to note that there is (at least  in  a  sufficiently
good approximation) \it no interaction \rm between subsystems 1
and 2 in this time interval. Besides, in good approximation, our
composite system is isolated.
\par We have described a preparator in a thought experiment. It might  be
real hard to construct a laboratory detector that does not
interact electromagnetically. This gives sufficient motivation to
discuss another modification of the standard S-G measuring
apparatus.

\subsection{The Second Modification for a S-G Preparator}

\indent The upper plate is removed again, but it is  not  replaced  by
 anything. The particle that would hit the upper plate in the standard
S-G instrument may now fly out freely. The lower plate is also
removed, and it is replaced by a particle detector that may be as
realistic as one wishes.
\par We want a so-called \it negative measurement \rm : it consists in the
anticoincidence of arrival of the particle on the plates and of
nondetection in the place of the lower plate. This amounts to \it
ideal occurrence \rm of $P^{(\bar n )}=P^+_1$, which is our
triggering event.
\par The described anticoincidence is hard to achieve in  the  laboratory
because it is not easy to make certain when the particle is
supposed to arrive at the plates. There is motivation for a third
modification.

\subsection{The Third Modification for a S-G Preparator}

\indent \rm We remove the upper plate, but we do not care about the lower plate.
We have a geometry that makes it possible to make us interested
only in the upper halfspace, where we put our measuring apparatus.
If it measures anything on the particle (at $t_f$ ) and one
obtains a result, then, due to the geometry, the particle must be
in the upper halfspace. Therefore, it must be in the state
$$ \Big( U_1(t_f-t_i,t_i)|\psi ^+ \rangle _1 \Big) |+,z \rangle _2 , $$
where $"U_1(...)"$ is a purely spatial evolution operator (the
spin does not change). This amounts to the same as if we had
occurrence of the triggering event $(P^+_1\otimes 1)$ at $t_i$ in
the state $|\Phi \rangle _{12}$ (and subsequent evolution).
\par To check if we are dealing with sufficiently general basic concepts
of preparation in standard quantum mechanics, let us take another,
a quite different and very well known example.

\section{PREPARATION THROUGH A HOLE}

\indent \rm Letting a beam through two successive holes in two parallel screens,
the preparation of a rather concentrated (e.g. pencil-shaped) beam
is achieved. This procedure consists of two equal stages. We start
by describing just one of them with the purpose to show that it
fits into the theoretical scheme suggested by the preceding
examples of preparation.

\par    In one-hole preparation the first subsystem is the screen,
the second is the particle. Let the screen be in the pure state
$|\psi ,t_i-\epsilon \rangle _2$ immediately before the initial moment of
the experiment $(0<\epsilon \ll 1)$. We think of the screen as of
an infinite surface perpendicular to the motion of the incoming
particle. Let the latter be in the pure state $|\chi ,t_i-\epsilon \rangle _2$.

\par    The screen is thought of as in some way broken up  into
nonoverlapping segments (the slit is one of them) enumerated by
"n" ($"\bar n "$ refers to the slit). Hitting the n-th segment,
i.e., transfer of linear momentum at this segment, corresponds to
the occurrence of the projector $P_1^{(n)}$. These (orthogonal)
projectors are assumed to add up as follows:
$$     \sum _nP_1^{(n)}=P_1 , \eqno{(5)} $$
and occurrence of the complementary projector $P_1^{\perp } \quad \Big( \equiv
 (1-P_1) \Big) $ has the physical meaning that the screen is not hit
at all (the particle has not reached it yet).

\par Correspondingly, the occurrence of the  particle  event
(projector) $Q_2^{(n)}$ means that the particle has hit the n-th
segment at $t_i$. Again $\sum_nQ_2^{(n)}\equiv Q_2$, and
$Q_2^{\perp }$ corresponds to the event that the particle has not
reached the screen yet. The composite-system state vector is
$$|\Phi ,t_i \rangle _{12}=\sum _n \Big[ \Big( P_1^{(n)}|\psi ,t_i \rangle _1 \Big) \Big( Q_2^{(n)}|\chi ,t_i \rangle _2 \Big) \Big]
 + \Big( P_1^{\perp }|\psi ,t_i \rangle _1 \Big) \Big( Q_2^{\perp }|\chi ,t_i \rangle _2 \Big) \eqno{(6)}$$
in full analogy with relation (1) for the S-G device. Here we
have, of course, assumed that the events occur in an ideal way,
i.e., that the L\"{u}ders state-projection formulae can be
applied. This is an oversimplification. (It will be improved upon
below.)

\par The second crucial entity  for  the  theory  of  preparation
is an observable $A_1$ with the spectral form
$$   A_1 \equiv \sum_na_nP_1^{(n)} + a^{\perp }P_1^{\perp },
\eqno{(7)} $$
where all characteristic values are
distinct(otherwise arbitrary but fixed). The \it triggering event
$P_1^{(\bar n )}$ \rm corresponds to the hole, and,
finally, the state of the particle at the final moment of
preparation is $cQ_2^{(\bar n )}|\chi ,t_i \rangle $, where "c" is a
normalization constant.

\par  In a \it more realistic theory, \rm the correlated
composite-system state is still a pure one, given by a state
vector that we decompose as follows

$$ |\Phi ,t_i \rangle _{12}\equiv \Big( P_1^{\perp }\otimes 1 \Big) |\Phi ,t_i \rangle _{12} +
   \sum_n \Big( P_1^{(n)}\otimes 1 \Big) |\Phi ,t_i \rangle _{12}. \eqno{(8)} $$
The second and third basic entities (i.e., $A_1$ and $P_1^{(\bar n )}$) are
unchanged, but the fourth, the state of the particle when the
preparation is completed, is (as known but not widely known):

$$ \rho _2^{(\bar n )}(t_i)\equiv Tr_1 \Big[ \Big( c(P_1^{(\bar n )}\otimes 1)
|\Phi ,t_i \rangle _{12} \Big) \Big(  \langle \Phi ,t_i|_{12}(P_1^{(\bar n )}\otimes 1)c \Big) \Big] ,
\eqno{(9)} $$ where "c" is a normalization constant,  and  $"Tr_1"$
denotes the partial trace over subsystem 1. (It is, of course,
assumed that $\Big( P_1^{(\bar n )}\otimes 1 \Big) |\Phi ,t_i \rangle _{12}\not=0$,
i.e., that the process considered allows passage through the hole
with positive probability.)

\par The state $\rho _2^{(\bar n )}$ is determined by the
composite-system state and the triggering event in the same way as
in the case of the S-G device (cf (4d) and (1)) or the
oversimplified composite-system state vector (6) for the one-hole
preparation. (We discuss in detail this partial-trace evaluation
below in Section 4.)

\par If one has  two  successive  holes,  as  one  usually  does
in the laboratory, then, denoting by $t'_i$ and $t_i$ the instants
of possible passage of the particle through the first and the
second hole respectively $(t'_i<t_i)$, the incoming state of the
particle in relation to the second hole is then

$$U_2(t_i-t'_i,t'_i) \, \rho _2^{(\bar n )}(t'_i) \, U_2^{\dagger }(t_i-t'_i,t'_i),$$
where $U_2$ is the evolution operator of the motion of the
particle between the two screens, and $\rho _2^{(\bar n )}(t'_i)$
is the state of the particle at $t'_i$. Then the described theory
is (essentially) repeated (in terms of mixed states).

\par In analogy with our above described modifications of the S-G device,
we can think of modifications of the hole-preparator.

\par \it The first one \rm goes as follows.

\par It is conceivable, in a thought experiment, to put a  detector
into the very hole. It should let the particles through without
changing their state and give information on the event of passage.

\par Naturally, passage results in the particle state
$\rho _2^{(\bar n )}(t_i)$ given by (9).

\par \it The  second modification \rm is achieved in the following way.

\par  We can imagine (in a thought experiment) all segments
of the screen being actually detectors except the hole itself. The
occurrence of the triggering event $P_1^{(\bar n )}$ amounts then
to the anticoincidence of the arrival of the particle to the
screen (nonoccurrence of $P_1^{\perp }$) and nonoccurrence of all
the events $\{ P_1^{(n)}:n\not= \bar n \}$. The prepared state of the
particle is again given by (9).

\par \it The third modification \rm  is, actually,
the original arrangement, when the state given by (9) is a
conditional one. It is valid if the particle passes the hole (but
we do not know that this or the opposite event occurs). Here the
geometry is trivially such that if anything is measured on the
particle to the right of the screen at $t_f$, the former must have
passed the hole, i.e., it is as if the triggering event had
occurred at $t_i$. (This will be discussed in detail in Section
5.)

\par We have now sufficient inductive  insight  for  a  general  standard
quantum mechanical theory of preparation. Nevertheless, it is
desirable to discuss further two points.

\par (i) As it was stated, the conditional state expressed by the
partial-trace formula (9) is known. But since it is not only
essential for preparation, but also for a further development of
the relative-collapse interpretation, we present its derivation in
the next section.

\par (ii) The precise meaning of the words "as if  the  triggering  event
occurred at $t_i$" in the case of $\rho _2^{(\bar n )}$ being a
conditional state (the third modification) must be explained in
more detail (in Section 5), the more so since it is based on a
recent result of the author.

\section{WHY THE PARTIAL-TRACE EVALUATION?}

\indent \rm Let $\rho _{12}$ be an arbitrary given composite-system
(mixed or pure) state (a statistical operator). Let, further,
$P_1$ be a first-subsystem event (projector) and let it occur in
whatever way in the state $\rho _{12}$. We want an answer to the
question: In what state leaves this occurrence the second
subsystem?

\par The sought-for state (statistical  operator) $\rho _2$
gives probability prediction for an arbitrary second-subsystem
event (projector) $Q_2$ through the formula $Tr_2 \Big( \rho _2Q_2 \Big) $, and,
as well-known, $\rho _2$ is thus determined by the totality of all
possible $Q_2$.

\par Since $P_1$ and $Q_2$ are compatible events (commuting projectors),
their coincidence on the one hand and the occurrence of $Q_2$ \it
immediately after \rm that of $P_1$ on the other is one and the
same thing. The coincidence probability can, as easily seen, be
written in a factorized form
$$ Tr_{12} \Big[ \rho _{12}(P_1\otimes Q_2) \Big] = \Big[ Tr_1 \Big( \rho _1P_1 \Big) \Big] \Big[ Tr_2 \Big( \rho _2Q_2 \Big) \Big], \eqno{(10)}$$
where $\rho _1$ is the state (reduced statistical operator)
of the first subsystem, $\rho _1\equiv Tr_2\rho _{12}$. The first
factor on the RHS is the probability of the event $P_1$ in $\rho _{12}$,
and, finally, $\rho _2$ is given by

$$ \rho _2\equiv \Big[ Tr_1 \Big( \rho _1P_1 \Big) \Big] ^{-1}Tr_1 \Big[ \rho _{12}(P_1\otimes 1) \Big] .
 \eqno{(11)} $$

\par Coincidence can be thought of as taking place in one
measurement, hence (10) can be viewed classically, and, the second
factor on the RHS of (10) is then, by definition, the
\it conditional probability \rm of $Q_2$ in the state in which the
second subsystem is left (immediately) after the occurrence of
$P_1$ in $\rho _{12}$. Since $\rho _2$ defined by (11) is a
statistical operator, as easily seen, and since $Q_2$ is an
arbitrary event, $\rho _2$ actually describes the mentioned state.
Hence, it is the sought-for expression justifying our
partial-trace evaluation in (9).

\section{THE CONDITIONAL STATE AND RETROACTIVE APPARENT
IDEAL OCCURRENCE}

\indent \rm When there is no detector in the preparator , i.e., when  it  is  no
measurement at all (the third modification in our discussions
above), then the most important of the four entities, the state
$\rho _2^{(\bar n )}(t_i)$ given by (9), has the meaning of a
conditional state, assuming validity under the condition that the
triggering event $(P_1^{(\bar n )}\otimes 1)$ occurs in the
composite-system state $\rho _{12}$ at $t_i$.

\par There is no actual occurrence of any event until $t_f$,
when a measurement result is obtained. Then, owing to the geometry
of the experiment, this amounts to the same, as it was claimed
above, as if $(P_1^{(\bar n )}\otimes 1)$ had occurred in $\rho
_{12}$. This requires additional explanation.

\par If any measurement result is obtained on the particle at $t_f$,
this takes place in a certain spatial region V (e.g. to the right
of the screen if the particle approaches the screen before $t_i$
from the left). Hence, the mentioned result coincides with the
occurrence of the event $Q_2(V)$ by which we mean that the
particle is in the mentioned region V.

\par If the triggering event $(P_1^{(\bar n )}\otimes 1)$ occurs
in $\rho _{12}(t_i)$ (i.e., if the screen undergoes the change -
linear-momentum transfer - corresponding to the particle's passage
through the hole), then at $t_f$ the event $Q_2(V)$ is certain to
occur in the state $U_{12}\rho _{12}(t_i)U^{\dagger }_{12}$, where
$U_{12} \quad \Big( \equiv U_{12}(t_f-t_i,t_i) \Big) $ is the evolution operator of
the composite system describing its evolution from $t_i$ till
$t_f$. This means that the particle must reach the region V.
Moreover, if the triggering event does not occur, i. e., if the
opposite event $[1-(P_1^{(\bar n )}\otimes 1)]$ occurs, at $t_i$
in $\rho _{12}$, then $Q_2(V)$ will not occur, i.e., $[1-Q_2(V)]$
will occur, at $t_f$ in the state $U_{12}\rho _{12}(t_i)U^{\dagger
}_{12}$.

\par There is a $\mbox{\rm theorem}^{(3)}$ that says that on account of
the two mentioned implications one must have
$$ (1\otimes Q_2) \Big[ U_{12}\rho _{12}(t_i)U^{\dagger }_{12} \Big] (1\otimes Q_2) \Big/
Tr_{12}\Big\{ (1\otimes Q_2) \Big[ U_{12}\rho _{12}(t_i)U^{\dagger }_{12} \Big] \Big\} =$$

$$ U_{12}\Big\{ \Big( P_1^{(\bar n )}\otimes 1\Big) \rho _{12}(t_i) \Big( P_1^{(\bar n )}\otimes 1 \Big) \Big/
Tr_{12} \Big[ \Big( P_1^{(\bar n )}\otimes 1 \Big) \rho _{12}(t_i) \Big] \Big\} U^{\dagger }_{12}.
\eqno{(12)} $$

\par This means that one has \it the same state \rm if, on the one hand, the
event $(1\otimes Q_2)$ occurs ideally at $t_f$  in the state
$[U_{12}\rho _{12}(t_i)U^{\dagger }_{12}]$, and,  on the other hand,
if the triggering event $(P_1^{(\bar n )}\otimes 1)$ occurs ideally
at $t_i$ in the state $\rho _{12}(t_i)$ and then the system evolves
till $t_f$.

\par If one utilizes the RHS of (12) instead of its LHS (for the composite-
system state at $t_f$), then one says that one has \it retroactive  apparent
ideal occurrence \rm (RAIO) of the event $(P_1^{(\bar n )}\otimes 1)$ in
$\rho _{12}(t_i)$ at $t_i$  as a  consequence of the \it actual occurrence \rm
of the event $(1\otimes Q_2)$ in $U_{12}\rho _{12}(t_i)U^{\dagger }_{12}$
at $t_f$.

\par Returning to our investigation, we consider the RAIO of
$(P_1^{(\bar n )}\otimes 1)$  in $\rho _{12}(t_i)$. We are actually
not interested in  the  composite-system  state, but in the state
of the second subsystem, i.e., of the particle. It is in the state
described by the reduced statistical operator:

$$ Tr_1 \Big[ \Big( P_1^{(\bar n )}\otimes 1 \Big) \rho _{12}(t_i) \Big( P_1^{(\bar n )}\otimes 1 \Big) \Big] \Big/
Tr_{12} \Big[ \Big( P_1^{(\bar n )}\otimes 1 \Big) \rho _{12}(t_i) \Big] , $$

\noindent which equals $\rho _2^{(\bar n )}(t_i)$ given by (9)
if one puts
$$ \rho _{12}(t_i)\equiv |\Phi ,t_i \rangle \langle \Phi ,t_i|. $$

\par Utilizing the identity

$$ 1 = (P_1^{(\bar n )}\otimes 1)+(P_1^{(\bar n )\perp }\otimes 1) , $$

\noindent we can write

$$ U_{12}|\Phi ,t_i \rangle =U_{12} \Big[ (P_1^{(\bar n )}\otimes 1)+
(P_1^{(\bar n )\perp }\otimes 1) \Big] |\Phi ,t_i \rangle _{12}. $$

\par Further, since the screen and the particle do not interact any  longer
if the latter has passed the hole, the evolution operator $U_{12}$
factorizes tensorically into the evolution operator of the screen $U_1$
and  that of the particle $U_2$ in this case (analogously as in
Subsection 2.3).  More precisely,

$$ U_{12} \Big( P_1^{(\bar n )}\otimes 1 \Big) = \Big( U_1\otimes U_2 \Big)
 \Big( P_1^{(\bar n )}\otimes 1 \Big) , \eqno{(13)} $$

\noindent and, owing to this, we can derive a simple form of the state
of the  particle at $t_f$  in the region V (relation (14) below).

\par Some event (corresponding to the  measurement  result)  is  going  to
occur in the region V. Since  the  measurement  apparatus  is  in  V,  the
occurrence of this event  implies  the  occurrence  of  the  event $Q_2(V)$.
Naturally, the measured event and $Q_2(V)$ are  compatible. We may imagine
that it is $Q_2(V)$ that occurs first, and the other event occurs immediately
afterwards. Since $Q_2(V)$ is not actually measured (only  implied),  we  are
justified to assume that its occurrence takes  place  in  the  ideal  way.
Hence, we take the LHS of (12) as the relevant composite-system state, and
we replace it by the RHS of (12). Then we obtain:

$$ \rho _2^{(\bar n )}(t_f)= $$

$$ Tr_1 \Big\{ U_{12} \Big[ \Big( P_1^{(\bar n )}\otimes 1 \Big) \rho _{12}
(t_i) \Big( P_1^{(\bar n )}\otimes 1 \Big) \Big/
Tr_{12} \Big( (P_1^{(\bar n )}\otimes 1)\rho _{12}(t_i) \Big) \Big] U_{12}^{\dagger } \Big\} . $$

\par Replacing here (13), we can take $U_2$  and $U_2^{\dagger }$
outside the partial  trace and we can omit $U_1$ and $U_1^{\dagger }$
under  the  partial  trace.  (These  are  known partial-trace identities.)
We finally obtain:

$$ \rho _2^{(\bar n )}(t_f)=U_2\rho _2^{(\bar n )}(t_i)U_2^{\dagger },
\eqno{(14)}$$
\vspace*{2mm}
$$ \rho _2^{(\bar n)}(t_i) \equiv Tr_1 \Big[ \Big( P_1^{(\bar n)} \otimes 1 \Big) \rho _{12}(t_i) \Big(P_1^{(\bar n)} \otimes 1 \Big) \Big/ Tr_{12} \Big( (P_1^{(\bar n)} \otimes 1) \rho _{12}(t_i) \Big) \Big]. \eqno{(15)} $$

\par It is perhaps worth noticing that also the event
$(1\otimes Q_2^{(\bar n )})$ of the
very passage of the particle through
the hole at $t_i$ may  play  a  role in
the theory. Actually, it stands in the
same relation to $(1\otimes Q_2(V))$ as
$(P_1^{(\bar n )}\otimes 1)$ does (cf
the above mentioned theorem and (12)).
Moreover, it stands in this same
relation also to the latter event
because $(1\otimes Q_2^{(\bar n )})$
occurs if and only if $(P_1^{(\bar n
)}\otimes 1)$ does. This makes these
two events $\mbox{\rm twin}^{(3)}$
events in the composite-system state
$\rho _{12}(t_i)$. This means that  the
two events  have also the same
probability in this state, and that
their  ideal occurrence changes the
state equally.

\section{GENERAL THEORY OF THE QUANTUM PREPARATOR
IN THE CONVENTIONAL INTERPRETATION OF QUANTUM MECHANICS}

\indent \rm Actually, both the S-G preparator and the hole preparator were
presented in the conventional  interpretation  of  quantum  mechanics (QM).
This is the text-book interpretation, actually a simplified form  of  the
so-called Copenhagen $\mbox{\rm one}^{(4)}$.

\par We now outline the general theory, and subsequently we point out some
unusual features in the conventional interpretation.

\par We call subsystem 2 the quantum object the state of which  is  going
to be prepared. Subsystem 1 is the preparator. There are two instants  $t_i$
and $t_f,\enspace t_i<t_f$. The former is the one when the  preparation  is
completed and the experiment begins. The latter is the final moment
of the  experiment, when some observable is measured on subsystem 2.

\par The two subsystems  interact  and  reach  a  composite-system  state
$\rho _{12}(t_i)$. This is the first basic entity of the  preparation.
The  second one is a first-subsystem observable $A_1$ with at least one
discrete characteristic projector $P_1^{(\bar n )}$, which is called
the triggering event, and  which represents the third basic entity
of preparation. The fourth basic entity is the conditional state
$\rho _2^{(\bar n )}(t_i)$ of the second subsystem  to  which  the
occurrence of the triggering event $P_1^{(\bar n )}$ in the state
$\rho _{12}(t_i)$ gives  rise. (The occurrence may take place in whatever
way, i.e., it need not be ideal) . The conditional state is  given  by  (15).

\par Finally, there is a fifth entity that belongs more to the experiment
than to the preparation. It is the evolution  operator
$$U_{12}\equiv U_{12}(t_f-t_i,t_i)$$
\noindent with the important property (13),
which means noninteraction between  object and preparator in the interval
from $t_i$ till $t_f$  after  the  triggering event has happened
(actually or retroactively apparently).

\par As a matter of fact, for a  given  preparator  it  is  only $U_2$,  the
evolution operator of the object, that must be known (cf (14)). As to $U_1$,
the evolution operator of the preparator, it is sufficient to know that it
exists, and that it enters the theory via (13). The concrete form of $U_1$ is
of no consequence.

\par In the conventional interpretation, we must distinguish two kinds of
preparators: the \it immediate-occurrence \rm or first-kind ones,
in which  the
triggering event does actually occur at $t_i$, and  the \it delayed-occurrence
\rm or second-kind ones, in which a special geometry singles out a spatial
region V, and some event (corresponding  to  obtaining  some  measurement
result on subsystem 2) actually occurs at the delayed moment $t_f$. (It  is
delayed as far as the preparation, not  the  experiment,  is  concerned).
But, as explained in detail in the preceding section, this gives rise  to
the retroactive apparent ideal occurrence (or  RAIO)  of  the  triggering
event, and the entire theory has exactly the same form as  for  a  first-
kind preparator.

\par  Now, I would like to point out some \it peculiar points \rm
in the  theory
in the conventional interpretation.

\par (i) We have one and the same formalism, but two  different  physical
mechanisms, i.e., the two kinds of preparators are equally described, but
we understand them as different processes.

\par (ii) The concept of RAIO, which enables us to describe both kinds of
preparators by the same formalism, is a paradoxical one:

\par In our hole-in-the-screen example, intuitively we do feel  that  the
particle must have passed the hole if it reaches region V. But  QM  seems
to prove us wrong. Since there is no measurement at $t_i$, there is no
collapse and no occurrence at that moment in actuality. The composite system
is decribed by $U_{12}\rho_{12}(t_i)U_{12}^{\dagger }$ at $t_f$.

\par This state includes, possibly in  a  coherent  (i.e.,  interference-allowing) way, also the possibility that the hole is not  passed  in  the
described example. At the final moment $t_f$, and only then, something
happens, some measurement result is obtained. From the very fact  that  this
result is obtained in the region V, we have the collapse described by the
LHS of (12). It does imply the RAIO of the triggering event, but this  is
only formal (or apparent) if we take occurrence (collapse)  really
seriously (as we should in the conventional interpretation).

\par In the conventional interpretation of QM one does not search for the
mechanism of the collapse, given rise to by the occurrence of some event.
But, the collapse is taken very seriously: it is considered to be a real,
objectively happening physical process.

\par  At this point we have the basic branching of interpretations of  QM:
the conventional one  with  numerous  foundational  attempts  to  explain
collapse with the help of one or another extra-quantum-mechanical agency,
and the \it no-collapse  approaches, \rm which  started  with the  theory
of $\mbox{\rm Everett}^{(5)}$. More will be said about them in the next section.

\par It is very important to realize that the collapse-a-real-process and
the no-collapse interpretations \it experimentally contradict each other. \rm
Thus, only one of them is actually true. Unfortunately,  the  experiments
required must find an observable incompatible with the pointer observable
on the (classical) measuring apparatus (cf  end  of  Section  6  in  Ref.
6). This has not succeeded so far.

\section{GENERAL THEORY OF THE  QUANTUM  PREPARATOR IN  THE
RELATIVE-COLLAPSE INTERPRETATION OF QUANTUM MECHANICS}

\indent \rm As it was mentioned, the no-collapse approach to QM started with the
famous article by $\mbox{\rm Everett}^{(5)}$. Nowadays, it receives far less attention
than the conventional approach. (Even its great initial admirer and, perhaps,
initiator, $\mbox{\rm Wheeler}^{(7),(8)}$ seems to have abandoned it.)

\par The Everett interpretation appears ambiguous to me: either it stipulates
collapse as a real process (it is called falling into a  branch  of
the universe), then it actually belongs  to  the  collapse-a-real-process
approaches, and is isomorphic to the conventional interpretation,  or  it
does not. But then it is unclear how  definite  measurement  results,  on
which every interpretation of QM hinges, come about.

\par To my mind, two elaborated and improved  forms  of  the  no-collapse
approach are the $\mbox{\rm modal}^{(9)}$ and the
$\mbox{\rm relative-collapse}^{(6)}$
interpretations. (See the discussion in Section 6 of Ref. 6.)
I'll utilize only the latter in relation to the theory of a
quantum preparator.

\par The relative-collapse (RC) interpretation of QM takes the idea of no
collapse seriously. There is no collapse either in first-kind or
in  second-kind preparation. Every measurement at $t_f$ is performed
in the composite system state
$\rho _{12}(t_f)=U_{12}\rho _{12}(t_i)U_{12}^{\dagger }$, which,
in  a  possibly  coherent way, contains both possibilities:
the case when the preparation did  succeed and that when it did not.

\par Nevertheless, the expounded quantum theory of  a  preparator  has  a
very simple form in this approach. And \it it need not distinguish two kinds
of preparators. \rm

\par In the RC interpretation of QM it  is  important  to  have  a
well-defined subject. This is given in terms of a  subsystem  observable.
The second basic entity $A_1$ of the expounded formalism of the preparator
suits this purpose perfectly.

\par One makes a shift of the cut (/) between subject (S) and object (O).
In terms of the so-called splits (that, by definition, encompass  besides
the cut also the subject and the object), the shift is  written  as  follows:

$$ S/O \equiv .../(1+2) \quad \rightarrow \quad S/O\equiv 1/2. $$

\par This means that in the first split the object (of  description)  was
the composite (1+2)-system, and one had an ill-defined subject. This  was
the case all the time so far. Then we join subsystem 1  (the  preparator)
to the subject. (Actually, the rest of the subject can  be  disregarded.)
Only subsystem 2 remains the object.

\par For every individual system one characteristic event of $A_1$,
in general called the subject event, in this case it is the triggering  event
$P_1^{(\bar n )}$, constitutes the "zero point" of the "coordinate system"
so to speak. The object is described by the conditional  state
$\rho _2^{(\bar n )}(t_i)$ that is determined by the condition
that we take the triggering event  as  occurring. Finally, we then we have
the evolution into the state

$$ \rho _2(t_f)=U_2 \, \rho _2^{(\bar n )}(t_i) \, U_2^{\dagger }. $$

\par Since there is no occurrence (collapse) as a real  physical  process
in this approach, it  is  of  no  consequence  if  we  have  "actual"  or
"retroactive apparent" occurrence of the triggering event at $t_i$. It is
anyway only our optimal choice of subject, i.e., how to get  rid  of  the
irrelevant in $\rho _{12}(t_i)$.

\par Just like in the theory of $\mbox{\rm measurement}^{(6)}$,
in the theory of preparators the RC interpretation of QM appears
to be best adapted
to  the  very formalism of QM. The fact that makes it hard to accept
is the  idea  that no events are occurring. Instead, only
the quantum  correlations  between
the subsystems are changing due to the interaction. The choice of a split
with a well-defined subject is only our subjective  way  of  reading  the
quantum correlations that are objectively there.

\par  Thus, the interaction between preparator and object brings about the
objectively existing quantum correlations in $\rho _{12}(t_i)$.
The described choice of subject brings out the relevant part,
the  object  state  $\rho _2^{(\bar n )}(t_i)$. And this is precisely
what is meant by preparation of the  state  of  the object.

\end{document}